\documentstyle[preprint,aps]{revtex}

\tightenlines 
\begin{document}

\draft
\preprint{AS-ITP-98-12, hep-ph/9810491; To appear in Phys. Rev. D}  
\title{SO(3) Gauge Symmetry  \\ and Neutrino-Lepton Flavor Physics }
\author{ Yue-Liang  Wu  }
\address{Institute of Theoretical Physics, Academia Sinica, \\
 P.O. Box 2735, Beijing 100080, P.R. China } 
\maketitle

\begin{abstract} 
 Based on the SO(3) gauge symmetry for three family leptons and 
general see-saw mechanism, we present a simple scheme that allows  
three nearly degenerate Majorana neutrino masses 
needed for hot dark matter. The vacuum structure of the spontaneous 
SO(3) symmetry breaking can automatically lead to a maximal CP-violating 
phase. Thus the current neutrino data on both the atmospheric 
neutrino anomaly and solar neutrino deficit can be accounted for 
via maximal mixings without conflict with the current data on 
the neutrinoless double beta decay. The model also allows rich interesting 
phenomena on lepton flavor violations.
\end{abstract}

       Studies on neutrino physics have resulted in the following 
 observations: i), The Super-Kamiokande data\cite{SUPERK1} on 
 atmospheric neutrino anomaly provide a strong evidence that 
 neutrinos are massive; ii), The current Super-Kamiokande data on 
 solar neutrino\cite{SUPERK2} cannot decisively establish 
  whether the solar neutrino deficit results from `just so' 
  oscillations\cite{JUSTSO} or MSW solutions\cite{MSW} with 
  small/large mixing angles\cite{BKS}. iii), To describe 
  all the neutrino phenomena such as the atmospheric neutrino anomaly, 
  the solar neutrino deficit and the results from the LSND experiment, 
  it is necessary to introduce a sterile neutrino\cite{STERILE}. 
  It indicates that with only three light neutrinos, one of the 
  experimental data must be modified; iv), The current experimental data 
  cannot establish whether neutrinos are Dirac-type or Majorana-type. 
  The failure of detecting neutrinoless double beta decay only provides, 
  for Majorana-type neutrinos, an upper bound on an `effective' 
  electron neutrino mass; v), Large neutrino masses around several 
  electron volts may play an important role in the evolution of 
  the large-scale structure of the universe\cite{MDM}. 
  Having noticed these facts, we will present in this 
 paper a conservative consideration in which only three light neutrinos 
are involved.

  To consider massive neutrinos, it is necessary to extend the minimal 
standard model (SM). The greatest success of the SM is its gauge symmetry
structure $SU_{L}(2)\times U_{Y}(1)$. 
As a simple extension of the standard model, 
it is of interest to introduce a flavor symmetry among 
the three families of the leptons. 
In this paper, we will choose the SO(3) symmetry group to describe
the lepton flavors and treat it as a gauge symmetry. 
As usual, we will also introduce an additional U(1) symmetry 
to construct the needed interaction terms. 
Let us begin with the following $SO(3)\times SU(2)_{L}\times U(1)_{Y}$ 
invariant lagrangian for leptons
\begin{eqnarray}
{\cal L} & = & {\cal L}_{YM} + \bar{L}_{i}i\gamma^{\mu}D_{\mu}L_{i} + 
\bar{e}_{R\ i}i\gamma^{\mu}D_{\mu}e_{R\ i} + \bar{N}_{R\ i}i\gamma^{\mu}
D_{\mu} N_{R\ i} + \bar{V}iD_{\mu}V + \bar{E}i\gamma^{\mu}D_{\mu} E  \nonumber \\
& + & y_{1}\bar{L}_{i}\varphi_{i} V + y_{2}\bar{V}\phi_{1} E + 
y_{3}\bar{E}\varphi_{i}e_{R\ i}+ y_{0}\bar{L}_{i}\phi_{2}N_{R\ i} 
 + S(x)\bar{N}_{R\ i} N_{R\ i}^{c} + H.c.    \\ 
&+ & M_{1} \bar{E}E + M_{2}\bar{V}V +  D^{\mu}\phi^{\ast}_{1}D_{\mu}\phi_{1} 
+  D^{\mu}\phi^{\ast}_{2}D_{\mu}\phi_{2} +  D^{\mu}\varphi^{\ast}D_{\mu}\varphi 
- V(\varphi_{i}, \phi_{1}, \phi_{2}, S) \nonumber 
\end{eqnarray}
with 
\begin{eqnarray}
D_{\mu}L_{i} & = & (\partial_{\mu}+i g_{1}\frac{1}{2}B_{\mu} 
-i g_{2}W_{\mu}^{k}\frac{\tau^{k}}{2})L_{i} 
-ig'_{3}A_{\mu}^{k}\frac{(t^{k})_{ij}}{2}L_{j} \nonumber \\
D_{\mu}e_{R\ i} & = & (\partial_{\mu}+ i g_{1}B_{\mu} )e_{R\ i}
-ig'_{3}A_{\mu}^{k}\frac{(t^{k})_{ij}}{2}e_{R\ j} \nonumber \\
D_{\mu}N_{R\ i} & = & \partial_{\mu} N_{R\ i}
-ig'_{3}A_{\mu}^{k}\frac{(t^{k})_{ij}}{2}N_{R\ j}  \nonumber \\
D_{\mu}V & = & (\partial_{\mu}-i g_{1}\frac{1}{2}B_{\mu} 
-i g_{2}W_{\mu}^{i}\frac{\tau^{i}}{2} )V  \\
D_{\mu}E & = & (\partial_{\mu}+i g_{1}B_{\mu} )E \nonumber \\
D_{\mu}\varphi_{i} & = & \partial_{\mu}\varphi_{i} -
ig'_{3}A_{\mu}^{k}\frac{(t^{k})_{ij}}{2} ) \varphi_{j} \nonumber  \\
D_{\mu}\phi_{1} & = & (\partial_{\mu}-i g_{1}\frac{1}{2}B_{\mu} 
-i g_{2}W_{\mu}^{k}\frac{\tau^{k}}{2})\phi_{1} \nonumber \\
 D_{\mu}\phi_{2} & = & (\partial_{\mu}+i g_{1}\frac{1}{2}B_{\mu} 
-i g_{2}W_{\mu}^{k}\frac{\tau^{k}}{2})\phi_{2} \nonumber
\end{eqnarray}
Where $i,j = 1,2,3$ being family indeces and $t^{k}$ ($k=1,2,3$) are
the three SO(3) generators. The U(1) charges of the various fields 
are chosen as: ($\varphi_{i}(x)$, $V(x)$, $E(x)$, $N_{R\ i}(x)$, 
$S(x)$, $e_{R\ i}(x)$, $L_{i}(x)$, $\phi_{1}(x)$, $\phi_{2}(x)$) = 
(-1/2, 1/2, 1/2, -1, -2, 1, 0, 0, 1), so that the theory
is anomaly free. $\bar{L}_{i}(x) = (\bar{\nu}_{i}, \bar{e}_{i})_{L}$ 
are the SU(2)$_{L}$ doublet leptons. $e_{R\ i}$ and $\nu_{R\ i}$ are 
right-handed charged leptons and neutrinos. $\phi_{1}(x)$ and 
$\phi_{2}(x)$ are the two Higgs doublets. $\varphi^{T}=
(\varphi_{1}(x), \varphi_{2}(x), \varphi_{3}(x))$ 
is a complex SO(3) triplet scalar and $S(x)$ is a complex 
singlet scalar. $V(x)$ is an SU(2)$_{L}$ doublet vector-like 
lepton with mass $M_{2}$ and $E(x)$ is a singlet vector-like 
electron with mass $M_{1}$. The three 
Majorana neutrinos $N_{R\ i}$ also belong to the SO(3) triplet 
representation. $A_{\mu} = A_{\mu}^{k}(x)t^{k}/2$ is the SO(3) 
gauge field and $g'_{3}$ is the SO(3) gauge coupling constant.
${\cal L}_{YM}$ represents the pure Yang-Mills gauge interactions and
$V(\varphi_{i}, \phi_{1}, \phi_{2}, S)$ denotes the Higgs potential.
If the singlet scalar S(x) carrys two unit 
lepton number, the lepton number is also conserved in the above 
lagarangian. The lepton number and U(1) symmetries are supposed to be 
spontaneously broken down at very high energy scales 
$<S(x)> = M_{N}\simeq 10^{13}$GeV and their effects will be not 
discussed in this paper. The mass scales $M_{1}$ and $M_{2}$ are 
assumed to be above the energy scales of the SO(3) flavor 
symmetry breaking which could 
occur at low energy scales. Adopting the general see-saw 
mechanism, we obtain the following effective lagrangian at the leading order
\begin{equation}
{\cal L} = C_{1}\frac{\varphi_{i}\varphi_{j}}{M_{1}M_{2}} \bar{L}_{i} \phi_{1}
 e_{R\ j} + C_{0}\frac{1}{M_{N}} \bar{L}_{i} \phi_{2}\phi_{2}^{T}L_{i}^{c} 
\end{equation}
with $C_{1}= y_{1}y_{2}y_{3}$ and $C_{0}=y_{0}^{2}$. This effective 
lagrangian remains invariant under $SO(3)\times 
SU_{L}(2)\times U_{Y}(1)$ action. After the symmetry is broken 
down to $U_{em}(1)$, one yields the following simple
 mass matrices for the Majorana neutrinos and charged leptons 
\begin{equation} 
M_{\nu}  = m_{0}\left( \begin{array}{ccc}
  1 & 0 & 0  \\
  0 & 1 & 0  \\
  0 & 0 & 1  \\
\end{array} \right)   
\end{equation}
and
\begin{equation}
M_{e} = \frac{m_{1}}{\sigma^{2}} \left( \begin{array}{ccc}
  \hat{\sigma}_{1}^{2} & \hat{\sigma}_{1}\hat{\sigma}_{2} & 
\hat{\sigma}_{1}\hat{\sigma}_{3}  \\
  \hat{\sigma}_{1}\hat{\sigma}_{2} & \hat{\sigma}_{2}^{2} & 
\hat{\sigma}_{2}\hat{\sigma}_{3}  \\
  \hat{\sigma}_{1}\hat{\sigma}_{3} & \hat{\sigma}_{2}
\hat{\sigma}_{3} & \hat{\sigma}_{3}^{3}  \\
\end{array} \right) 
\end{equation}   
with $m_{0} = C_{0}v_{2}^{2}/M_{N}$ and 
$m_{1} = C_{1}v_{1}\sigma^{2}/M_{1}M_{2}$. 
Where $v_{1,2} = <\phi_{1,2}>$, $\hat{\sigma}_{i} = <\varphi_{i}>$ 
and $\sigma = \sqrt{|\hat{\sigma}_{1}|^{2} + |\hat{\sigma}_{2}|^{2} + 
|\hat{\sigma}_{3}|^{2}}$ are the vacuum expectation values of 
the corresponding fields after spontaneous symmtry breaking. Here 
the values of the quantities $y_{2}v_{1}$ and $|\hat{\sigma}_{i}|$ 
are considered to be smaller than the mass scales $M_{1}$ and $M_{2}$, i.e., 
$y_{2}v_{1}, |\hat{\sigma}_{i}| < M_{1}, M_{2}$. 

   As the SO(3) flavor symmetry is treated as a gauge symmetry, 
one can always express the complex SO(3) triplet  
scalar fields $\varphi_{i}(x)$ in terms of the three rotational fields and three 
amplitude fields. This is analogous to SU(2) gauge symmetry, the complex SU(2) doublet 
scalar field can always be expressed in terms of three SU(2) `rotational'  fields and one 
amplitude field. As the SO(3) rotation matrix is real, which is unlike to the SU(2) rotation 
matrix that is complex, one of the three amplitude fields of the complex SO(3) triplet  
scalar  must be a pure imaginary field so that one can generate the complex SO(3) triplet scalar 
fields $\varphi_{i}(x)$ by an SO(3) field  $O(x)= e^{i\eta_{i}(x)t^{i}} \in $ SO(3) action on 
the three amplitude fields. Explicitly, one has 
\begin{equation} 
\left( \begin{array}{c}
  \varphi_{1}(x) \\
  \varphi_{2}(x)   \\
  \varphi_{3}(x)   \\
\end{array} \right) = e^{i\eta_{i}(x)t^{i}} \frac{1}{\sqrt{2}}
\left( \begin{array}{c}
  \rho_{1}(x) \\
  i\rho_{2}(x)   \\
  \rho_{3}(x)   \\
\end{array} \right)   
\end{equation}
where the three real rotational fields 
$\eta_{i}(x)$ and the three real amplitude fields $\rho_{i}(x)$ reparameterize the six real 
fields of the complex triplet scalar field $\varphi(x)$. In general there are three possible 
assignments for the imaginary amplitude. In this note, we only consider the case that  
the imaginary part is assigned to the second amplitude field. From the Higgs mechanism, 
the three rotational fields $\eta_{i}(x)$ will be `eaten' by the 
three gauge fields $A_{\mu}^{i}(x)$ after spontaneous symmetry breaking 
\begin{equation}
\rho_{i}(x)\rightarrow \sigma_{i} + \rho_{i}(x)
\end{equation}
For this case, one has $\hat{\sigma}_{1} = \sigma_{1}$, $\hat{\sigma}_{2} = i\sigma_{2}$ and 
$\hat{\sigma}_{3} = \sigma_{3}$. With this analysis, it is seen that when three amplitude
fields get nonzero vacuum expectation values, an SO(3) gauge theory with one complex triplet 
Higgs field always spontaneously breaks CP invariance via a $\pi/2$ phase. 

   It is easy to check that the matrix $M_{e}$ is a rank one matrix and 
 diagonalized by a unitary matrix $U_{e}$ via  
\begin{equation}
D_{e} = U_{e}^{\dagger} M_{e} U_{e}^{\ast}
\end{equation}
with
\begin{equation} 
D_{e}  = \left( \begin{array}{ccc}
  0 & 0 & 0  \\
  0 & 0 & 0  \\
  0 & 0 & m_{1}  \\
\end{array} \right)   
\end{equation}
and 
\begin{equation}
U_{e}=\left( \begin{array}{ccc}
  -ic_{1} & c_{2}s_{1} & s_{1} s_{2}  \\
 -s_{1}  & ic_{1}c_{2} & ic_{1}s_{2} \\
0 &  -s_{2}  & c_{2}  \\
\end{array} \right)
\end{equation}
where we have used the notations $s_{1} \equiv \sin \theta_{1} = \sigma_{1}/\sigma_{12}$, 
$s_{2} \equiv \sin \theta_{2} = \sigma_{12}/\sigma$, 
$\sigma_{12} = \sqrt{\sigma_{1}^{2} + \sigma_{2}^{2}}$ and $\sigma =
\sqrt{\sigma_{12}^{2} + \sigma_{3}^{2}}$.

As the neutrinos are in the mass eigenstate basis, the CKM-type lepton 
mixing matrix $U_{LEP}$ defined in the charged weak gauge interactions
\begin{equation}  
{\cal L}_{W}=\bar{e}_{L}\gamma^{\mu}U_{LEP} \nu_{L} W_{\mu}^{-} 
+ H.c.
\end{equation}
is simply given by 
\begin{equation} 
U_{LEP} =  U_{e}^{\dagger} =\left( \begin{array}{ccc}
  ic_{1} & -s_{1} & 0  \\
 c_{2}s_{1} & -ic_{1}c_{2} & -s_{2} \\
 s_{1} s_{2} & -ic_{1}s_{2} & c_{2}  \\
\end{array} \right)
\end{equation}
Before proceeding, we would like to address the following points: Firstly, as neutrino mass matrix 
is a Majorana one, the complex $U_{LEP}$ cannot be absorbed.  Secondly, as the neutrino masses
are degenerate at the leading order of the present considerations, the unitary matrix $U_{LEP}$ 
is arbitray up to multiplication by an arbitray orthogonal matrix from the right. 
This arbitrariness can be fixed by eliminating the degeneracy of the neutrino masses, this may be 
realized by considering additional new contributions and high order corrections 
to the neutrino masses. It will also be seen below that such an arbitrariness is actually prevented from
the SO(3) gauge interactions of the neutrinos. Thirdly, as the matrix element $(U_{LEP})_{13} = 0$ 
such a unitary matrix can always be transformed into an orthogonal matrix by a phase redefinition of 
the left-handed neutrinos and charged leptons. For the above case, it can be realized by redefining 
$e_{L}\rightarrow i e_{L}$ and $\nu_{\mu L} \rightarrow i \nu_{\mu L}$, one then has
\begin{equation} 
U_{LEP} \rightarrow O_{LEP} =\left( \begin{array}{ccc}
  c_{1} & -s_{1} & 0  \\
 c_{2}s_{1} & c_{1}c_{2} & -s_{2} \\
 s_{1} s_{2} & c_{1}s_{2} & c_{2}  \\
\end{array} \right)
\end{equation}
However, we would like to emphasize that such a phase redefinition is not trivial due to 
Majorana neutrinos and SO(3) gauge interactions. This is because the neutrino mass matrix becomes, 
after the phase redefinition $\nu_{\mu L} \rightarrow i \nu_{\mu L}$, the following form
\begin{equation} 
M_{\nu}\rightarrow M_{\nu}  = m_{0}\left( \begin{array}{ccc}
  1 & 0 & 0  \\
  0 & -1 & 0  \\
  0 & 0 & 1  \\
\end{array} \right)   
\end{equation}
which is no longer a unit matrix and the neutrinos are not in the physical mass basis. 
A direct physical effect of such an CP phase can be seen in the processes concerning 
the neutrinoless double beta decay. One can also explicitly see below that the phase redefinition 
will lead to a change of the phases of the couplings in the SO(3) gauge interactions of the charged 
leptons and neutrinos. 

  When transfering the neutrino mass basis into the weak gauge and charged lepton mass basis,
the neutrino mass matrix $M_{\nu}=m_{0} U_{LEP}U_{LEP}^{T}$ 
is found to have the following form 
\begin{equation}
M_{\nu}= m_{0}\left( \begin{array}{ccc}
  s_{1}^{2}-c_{1}^{2} & 2ic_{1}s_{1}c_{2} & 2ic_{1}s_{1}s_{2}  \\
   2ic_{1}s_{1}c_{2} & c_{2}^{2}(s_{1}^{2}-c_{1}^{2})+s_{2}^{2}
 & c_{2}s_{2}(s_{1}^{2}-c_{1}^{2}) - c_{2}s_{2}  \\
  2ic_{1}s_{1}s_{2} & c_{2}s_{2}(s_{1}^{2}-c_{1}^{2}) - c_{2}s_{2}
 & s_{2}^{2}(s_{1}^{2}-c_{1}^{2})+ c_{2}^{2}  \\ 
\end{array} \right)
\end{equation}
which can be diagonalized by a unitary matrix $U_{\nu}$ via 
$U_{\nu}^{T}M_{\nu}U_{\nu}$ with $U_{\nu}= U_{LEP}$. 

 To describe the realistic world, one needs to consider mechanisms that 
will split the degeneracy of the three neutrinos and produce the correct 
muon and electron masses. These mechanisms can arise from high order 
radiative corrections or suppressed contributions due to additional 
new interaction terms. As the mass differences among the three neutrinos are expected to be very 
small, the muon and electron masses are known to be much smaller than the tau mass,
it is then reasonable to obtain a realistic pattern which 
will not significantly differ from the current pattern. 
For simplicity of discussions, we shall leave such constructions 
elsewhere\cite{YLWU} and consider the present scheme, similar to some other schemes such as 
the democratic scheme, to be the first essential step. 

 As an approximate estimation, the current experimental data\cite{DBD} 
from neutrinoless double beta decay indicates that
\begin{equation}
 <m_{\nu_{e}}> = m_{0}(s_{1}^{2}-c_{1}^{2})< 0.46 eV
\end{equation}
The required hot dark matter needs $m_{0} \simeq 1.5$ eV. Combinations
of these two experimental data will result in a bound on $\theta_{1}$
\begin{equation}
36^{0}< \theta_{1} < 54^{0},\qquad or, \qquad \sin^{2}2\theta_{1} > 0.9
\end{equation}
which shows that when neutrinos do play an essential 
role for the evolution of large scale structure of the universe,
the solar neutrino deficit appears to be explained 
by the $\nu_{e}-\nu_{\mu}$ mixing via `Just So' oscillations or 
MSW solution with large mixing angle. Furthermore, the current 
Super-Kamiokande data on atmospheric neutrinos require 
\begin{equation}
\sin^{2}2\theta_{2} > 0.8,\qquad or,\qquad 32^{0} < \theta_{2} < 58^{0}
\end{equation} 

  These consequences motivate us to consider scenarios in which the
 two vacuum expectation values of $\sigma_{1}$ and $\sigma_{2}$ are 
equal, i.e., $\sigma_{1}=\sigma_{2}$ (or  $s_{1}=1/\sqrt{2}$), 
 the neutrino mass and mixing matrices can be simply written as   
\begin{equation}
M_{\nu} = m_{0}\left( \begin{array}{ccc}
  0 & ic_{2} & is_{2}  \\
   ic_{2} & s_{2}^{2}
 &  - c_{2}s_{2}  \\
  is_{2} &  - c_{2}s_{2}
 &  c_{2}^{2}  \\ 
\end{array} \right)
\end{equation}
and 
\begin{equation}
U_{LEP}= 
\left( \begin{array}{ccc}
  \frac{1}{\sqrt{2}}i & -\frac{1}{\sqrt{2}} & 0  \\
  \frac{1}{\sqrt{2}}c_{2} & -\frac{1}{\sqrt{2}}c_{2}i & -s_{2} \\
 \frac{1}{\sqrt{2}}s_{2} & -\frac{1}{\sqrt{2}}s_{2}i & c_{2}  \\
\end{array} \right)
\end{equation}
which arrives at the pattern suggested in\cite{FV}.

 Two particular interesting cases which have been widely discussed 
in the literature can be easily achieved by further considering the 
following two cases: First, $\sigma_{3}^{2} = \sigma_{1}^{2} + 
\sigma_{2}^{2}$ and $\sigma_{1} = \sigma_{2}$, 
namely, $s_{1}=s_{2}=1/\sqrt{2}$, 
one then arrives at the bi-maximal mixings\cite{BMAX,BMAXM} with 
a maximal CP-violating phase. Explicitly, the neutrino mass and mixing 
matrices read 
\begin{equation}
M_{\nu} = m_{0}\left( \begin{array}{ccc}
  0 & \frac{1}{\sqrt{2}}i & \frac{1}{\sqrt{2}}i  \\
   \frac{1}{\sqrt{2}}i & \frac{1}{2} &  - \frac{1}{2}  \\
  \frac{1}{\sqrt{2}}i &  - \frac{1}{2} &  \frac{1}{2}  \\ 
\end{array} \right)
\end{equation}
and 
\begin{equation}
U_{LEP}= 
\left( \begin{array}{ccc}
  \frac{1}{\sqrt{2}}i & -\frac{1}{\sqrt{2}} & 0  \\
  \frac{1}{2} & -\frac{1}{2}i & -\frac{1}{\sqrt{2}} \\
 \frac{1}{2} & -\frac{1}{2}i & \frac{1}{\sqrt{2}}  \\
\end{array} \right)
\end{equation}
which yields the pattern discussed recently 
by Georgi and Glashow\cite{GG}.

Second, when three vacuum expectation values $\sigma_{i}$ (i=1,2,3) 
are democratic, i.e., $\sigma_{3} = \sigma_{1} = \sigma_{2}$,
namely $s_{1}=1/\sqrt{2}$ and $s_{2}=\sqrt{2/3}$, we 
then obtain the democratic mixing\cite{DEM,DEMM} with a maximal
CP-violating phase. The explicit neutrino mass and mixing matrices 
are given by 
\begin{equation}
M_{\nu} = m_{0}\left( \begin{array}{ccc}
  0 & \frac{1}{\sqrt{3}}i & \frac{2}{\sqrt{6}}i  \\
   \frac{1}{\sqrt{3}}i & \frac{2}{3} &  - \frac{\sqrt{2}}{3}  \\
  \frac{2}{\sqrt{6}}i  &  - \frac{\sqrt{2}}{3} &  \frac{1}{3}  \\ 
\end{array} \right)
\end{equation}
and 
\begin{equation}
U_{LEP} = 
\left( \begin{array}{ccc}
  \frac{1}{\sqrt{2}}i & -\frac{1}{\sqrt{2}} & 0  \\
  \frac{1}{\sqrt{6}} & -\frac{1}{\sqrt{6}}i & - \frac{2}{\sqrt{6}} \\
 \frac{1}{\sqrt{3}} & -\frac{1}{\sqrt{3}}i & \frac{1}{\sqrt{3}}  \\
\end{array} \right)
\end{equation}
which comes to a similar form provided by Mohapatra\cite{RABI} for the 
degenerate neutrino scenario.  

 So far we have mainly focused on the neutrino masses and mixings. 
In fact, rich new interesting physical phenomena 
may arise from the SO(3) gauge interactions. We would like to address 
the following important point that the redifinition 
of the phases of the leptons is not trivial in the present model, 
which is unlike to the standard model. This is because the SO(3) 
gauge fields couple to lepton flavor changing neutral currents. 
Explicitly, the SO(3) gauge interactions in the mass eigenstate 
of the leptons have the following form
\begin{equation}
{\cal L}_{F} = \frac{1}{2} g'_{3} \bar{\nu}_{L}t^{i}\gamma^{\mu}\nu_{L}
\ A_{\mu}^{i} + \frac{1}{2}g'_{3} \bar{e}_{L} K_{e}^{i}\gamma^{\mu}e_{L}
\ A_{\mu}^{i} - \frac{1}{2}g'_{3}\bar{e}_{R} K_{e}^{i \ast}
\gamma^{\mu}e_{R}\ A_{\mu}^{i} 
\end{equation} 
with $K_{e}^{i}= U_{e}^{\dagger}t^{i}U_{e}$. Using the relation $K_{e}^{i}= 
U_{e}^{\dagger}t^{i}U_{e} = K^{ia}_{e}t^{a}$ with $t^{a}$ 
($a=1, \cdots, 9$) being the nine U(3) generators, the above interactions can be 
reexpressed as
\begin{equation}
{\cal L}_{F}  = \frac{1}{2}g'_{3} \bar{\nu}_{L}t^{i}\gamma^{\mu}\nu_{L}\ A_{\mu}^{i}  
+ \frac{1}{2}g'_{3} \bar{e}\bar{K}^{i}_{e}\gamma^{\mu}e\ A_{\mu}^{i} 
+ \frac{1}{2}g'_{3}\bar{e}\tilde{K}^{i}_{e}\gamma_{5}\gamma^{\mu}e\ A_{\mu}^{i} 
\end{equation}   
with $\bar{K}^{i}_{e} = K^{ij}_{e}t^{j}$ for vector currents 
and $\tilde{K}^{i}_{e}= K^{ia'}_{e}t^{a'}$ for axial vector currents. 
Where $t^{i}$ ($i=1,2,3$) are the three complex (or antisymmetric) generators 
of U(3) and $t^{a'}$ ( $a'= 4, \cdots, 9$) are the six real (or symmetric)
generators of U(3). Note that the above interaction forms are given 
at the SO(3) gauge basis. After spontaneous symmetry breaking of SO(3), 
the gauge fields $A_{\mu}^{i}$ receive masses by 'eating' the three rotational 
fields $\eta^{i}(x)$. For the SO(3) vacuum structure given above, 
$A_{\mu}^{1}$ and $A_{\mu}^{3}$ are not in the mass eigenstates since
they mix each other. Denoting the physical gauge fields as $F_{\mu}^{i}$, 
we have $A_{\mu}^{i} = O_{F}^{ij}F_{\mu}^{j}$. Explicitly, it reads 
\begin{equation}
 \left( \begin{array}{c} 
A_{\mu}^{1} \\ A_{\mu}^{2} \\ A_{\mu}^{3} \\
\end{array} \right)  = \left( \begin{array}{ccc}
  c_{3} & 0  & -s_{3} \\
   0 & 1 & 0  \\
s_{3}  &  0 & c_{3}  \\ 
\end{array} \right) \left( \begin{array}{c} 
F_{\mu}^{1} \\ F_{\mu}^{2} \\ F_{\mu}^{3} \\
\end{array} \right) 
\end{equation}   
with the mixing angle $s_{3} = \sin\theta_{3} = 
\sigma_{1}/\sigma_{13}$\ ($\sigma_{13} =\sqrt{\sigma_{1}^{2}+
 \sigma_{3}^{2}}$). Their physical masses are:
\begin{equation}
m_{F_{1}} = \frac{1}{2}g'_{3}\sigma_{2},\qquad
 m_{F_{2}} = \frac{1}{2}g'_{3}\sigma_{13}, 
\qquad m_{F_{3}} = \frac{1}{2}g'_{3}\sigma
\end{equation}
In the physical mass basis of leptons and gauge bosons, 
the above interactions will be 
\begin{equation}
{\cal L}_{F} =  \frac{1}{2}g'_{3} \bar{\nu}_{L}t^{j}O_{F}^{ji}\gamma^{\mu}\nu_{L}
\ F_{\mu}^{i} + \frac{1}{2}g'_{3} \bar{e}_{L} V_{e}^{i}\gamma^{\mu}e_{L}
\ F_{\mu}^{i} - \frac{1}{2}g'_{3}\bar{e}_{R} V_{e}^{i \ast}
\gamma^{\mu}e_{R}\ F_{\mu}^{i} 
\end{equation}
with $V_{e}^{i} = K_{e}^{j}O_{F}^{ji} = U_{e}^{\dagger}t^{j}U_{e}O_{F}^{ji}$. 
To be explicit, for the case considered above, we have 
\begin{equation}
K_{e}^{1} = \left( \begin{array}{ccc}
  2c_{1}s_{1} & ic_{2}(s_{1}^{2}-c_{1}^{2}) & is_{2} (s_{1}^{2}-c_{1}^{2}) \\
   -ic_{2}(s_{1}^{2}-c_{1}^{2}) & 2c_{1}s_{1}c_{2}^{2} &  2c_{1}s_{1} c_{2}s_{2}  \\
  -is_{2}(s_{1}^{2}-c_{1}^{2}) &  2c_{1}s_{1} c_{2}s_{2} & 2c_{1}s_{1}s_{2}^{2}  \\ 
\end{array} \right)
\end{equation} 
\begin{equation}
K_{e}^{2} = \left( \begin{array}{ccc}
  0 & c_{1}s_{2} & -c_{1}c_{2}  \\
   c_{1}s_{2} & 0 &  is_{1}  \\
 -c_{1}c_{2} &  -is_{1} & 0 \\ 
\end{array} \right)
\end{equation} 
and 
\begin{equation}
K_{e}^{3} = \left( \begin{array}{ccc}
  0 & is_{1}s_{2} & -is_{1}c_{2}  \\
   -is_{1}s_{2}  & 2c_{1}c_{2}s_{2} &  (s_{2}^{2}-c_{2}^{2})c_{1}  \\
 is_{1}c_{2} &  (s_{2}^{2}-c_{2}^{2})c_{1} & -2c_{1}c_{2}s_{2}  \\ 
\end{array} \right)
\end{equation} 
and 
\begin{eqnarray}
V_{e}^{1} & = & \cos\theta_{3}K_{e}^{1} + \sin\theta_{3}K_{e}^{3}, \nonumber \\
 V_{e}^{2} & = & K_{e}^{2}, \\
V_{e}^{3} & = & -\sin\theta_{3}K_{e}^{1} + \cos\theta_{3}K_{e}^{3} \nonumber 
\end{eqnarray} 
If the mixing matrix $U_{e}$ does not significantly affected when muon lepton gets 
its physical mass, the current data on lepton flavor violating process 
$\mu \rightarrow 3e$ with $Br(\mu \rightarrow 3e) < 1\times 10^{-12}$
\cite{LFV} lead to the following constraints on the SO(3)
symmetry breaking scale 
\begin{equation}
\sigma_{1} \sim \sigma_{2}\sim \sigma_{3} > 10^{3}\ v 
\end{equation}
with $v=246$ GeV being the weak symmetry breaking scale. 
 
   In conclusion, it has been shown that the SO(3) gauge
symmetry for lepton flavors appears to have some remarkable features which 
are applicable to the current interesting phenomena concerning
neutrinos. 

{\bf Acknowledgments:} This work was supported in part by the NSF of China under the grant No. 19625514.


\end{document}